\documentclass{aa}  
\usepackage{natbib}
\usepackage{graphicx}
\usepackage{txfonts}
\begin{document}
   \title{Thermal shielding of an emerging active region}
   %\subtitle{Thermal shielding as observed by SDO/AIA and SDO/HMI}

   \author{S. R\'egnier
          %\inst{1}
          %\and
          %R. W. Walsh
          }

   \offprints{S. R\'egnier}

   \institute{Jeremiah Horrocks Institute, University of Central Lancashire,
   Preston, Lancashire, PR1 2HE, UK\\
              \email{SRegnier@uclan.ac.uk}
         %\and
         %    University of Alexandria, Department of Geography, ...\\
         %    \email{c.ptolemy@hipparch.uheaven.space}
         %   \thanks{The university of heaven temporarily does not
         %            accept e-mails}
             }

   \date{Received ; accepted }

% \abstract{}{}{}{}{} 
% 5 {} token are mandatory
 
  \abstract
   {The interaction between emerging active regions and the pre-existing coronal
   magnetic field is important to understand better the mechanisms of storage
   and release of magnetic energy from the convection zone to the high corona.
   %Emergence of active region in the photosphere and its interaction with the
   %corona is an important process to understand better the mechanisms of energy
   %storage and release in the convection zone. 
   }
   {We are aiming at describing the first steps of the emergence of an active
   region within a pre-existing quiet-Sun corona in terms of the thermal and
   magnetic structure.}
   {We use unprecedented spatial, temporal and spectral coverage from the
   Atmospheric Imager Assembly (AIA) and from the Helioseismic and Magnetic
   Imager (HMI) on board the Solar Dynamics Observatory (SDO).}
   {Starting on 30 May 2010 at 17:00 UT and for 8 hours, we follow the
   emergence of the active region AR11076 within a quiet-Sun region. Using several
   SDO/AIA filters covering temperatures from 50000K to 10 MK, we show that the
   emerging process is characterised by a thermal shield at the interface
   between the emerging flux and pre-existing quiet-Sun corona.}
   {The active region 11076 can be considered as a peculiar example of emerging
   active region as (i) the polarities emerge in a photospheric quiet-Sun region
   near a supergranular-like distribution, (ii) the polarities forming the
   bipolar emerging structure do not rotate with respect to each other
   indicating a small amount of twist in the emerging flux bundle. There is a
   thermal shield formed at the interface between the emerging active region and
   the pre-existing quiet-Sun region. The thermal shielding structure
   deduced from all SDO/AIA channels exhibits a strong asymmetry between the two
   polarities of the active region suggesting that the heating mechanism for one
   polarity is more likely to be magnetic reconnection, whilst it is due to
   increasing magnetic pressure for the opposite polarity.}


   \maketitle

%%%%%%%%%%%%%%%%%%%%%%%%%%%%%%%%%%%%%%%%%%%%%%%%%%%%%%%%%%%%%%%%%%%%%%%%%%%%
\section{Introduction}
%%%%%%%%%%%%%%%%%%%%%%%%%%%%%%%%%%%%%%%%%%%%%%%%%%%%%%%%%%%%%%%%%%%%%%%%%%%%

% Intro to flux emergence

Flux emergence is a process transporting magnetic energy and plasma from
the convection zone into the solar atmosphere. The standard models of flux
emergence \citep[see reviews by][]{arc08,hoo11} are based (i) on the buoyancy of
twisted flux tubes generated near the tachocline by the global dynamo action, or
(ii) on the generation of small scale magnetic field in flux sheet regions in a
thin layer near the photosphere. As the magnetic flux is
transported by convective motions, emerging regions are most likely to appear at
the centre of supergranular cells where strong radial upflows and/or horizontal
flows are observed \citep{lei62}. The emerging flux is first observed on the
photosphere as newly formed, highly concentrated magnetic elements. The magnetic
elements are thus growing in size and moving apart to build magnetic polarities
such as pores and sunspots. Important features of the emergence of twisted flux
tubes are the increase of the total unsigned magnetic flux, the separation of
polarities when the flux is increasing, and the rotation of magnetic polarities
with respect to each other due to the transport of twist (or helicity) into the
corona \citep{zwa85,lop03}. Extensive observations of emerging active regions in
quiet-Sun areas or in already emerged active regions have already been reported
\citep[e.g.,][]{zwa85,van00a,oka08,par09,can09,har10,gom10,var11}. Despite these
observations, two parts of the flux emergence into the corona have not yet been
tackled in depth due to the lack of high cadence and wide temperature coverage
in terms of magnetic field and EUV emission: (i) the very first steps of the
emergence (the first few hours say) before pores are formed and the emerged flux
has just reached coronal heights, (ii) the steady and impulsive response of the
corona during that period. 

AR 11076 was observed to emerge on 30 May 2010 in the southern hemisphere. To
determine the initial time of the flux emergence, we use the capabilities of
SDO/HMI by combining both line-of-sight magnetic field and continuum images to
track down in time the first signature of the emergence. Hence we define the 30
May 2010 at 17:00 UT as the start of the emergence. We thus combine SDO/HMI and
SDO/AIA to study the interaction of the emerging magnetic field and the
pre-existing quiet-Sun corona. We restrict the study to the first eight
hours of emergence just before the formation of a pore. The sunspots (umbra and
penumbra) will form later on 31 May around 12:00 UT. We also give a physical
interpretation of the formation of a thermal shield.

%%%%%%%%%%%%%%%%%%%%%%%%%%%%%%%%%%%%%%%%%%%%%%%%%%%%%%%%%%%%%%%%%%%%%%%%%%%%
\section{Thermal and magnetic evolution from SDO/AIA and SDO/HMI}
\label{sec:aia_obs}
%%%%%%%%%%%%%%%%%%%%%%%%%%%%%%%%%%%%%%%%%%%%%%%%%%%%%%%%%%%%%%%%%%%%%%%%%%%%

% thermal evolution 
% images 171, 193, 211
% images  94 + magnetograms

We describe the evolution of the emerging active region 11076 during eight hours
between 17:00 UT on 30 May 2010 to 01:00 UT on 31 May 2010. We use both SDO/HMI
\citep{sch12} and SDO/AIA instruments to study the magnetic and thermal structure of the
emerging region. SDO/AIA data are level 1 images using a first approximation of
the calibration, which does not change the results reported in this letter. The
pixel size is 0.56\arcsec and the time cadence for this study is 45s for SDO/HMI
and 36s for SDO/AIA. SDO/AIA capabilities and thermal responses are described in
\citet{lem12} \citep[see also][]{odw10}. We thus define the temperature of a
channel as the temperature of the peak of emission as reported by \cite{lem12},
keeping in mind that all broad-band channels are multithermal.

\begin{figure}[t]
\centering
% mag
\includegraphics[width=.327\linewidth, bb=135 60 370 295, clip]
	{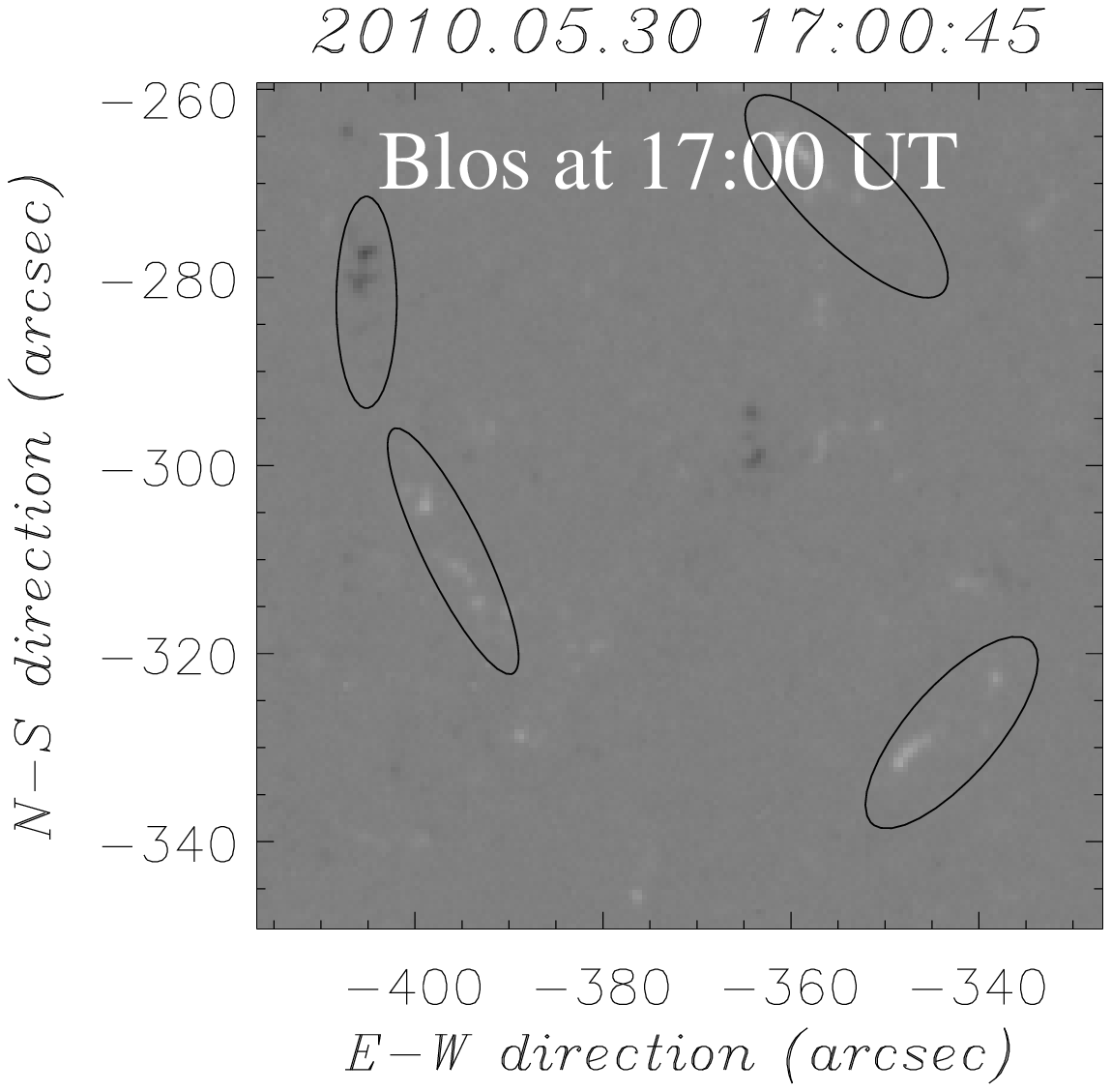}
\includegraphics[width=.327\linewidth, bb=135 60 370 295, clip]
	{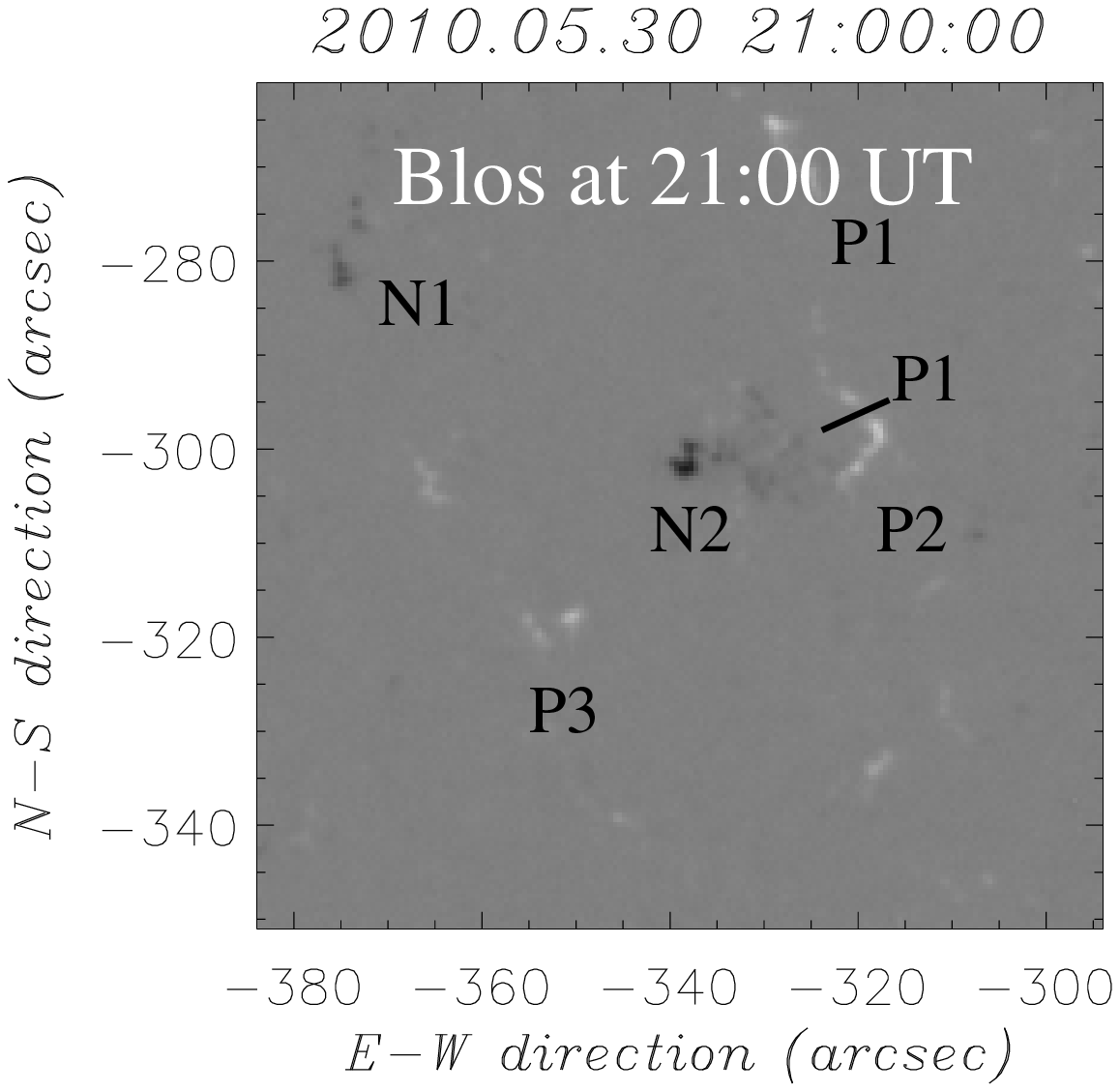}
\includegraphics[width=.327\linewidth, bb=135 60 370 295, clip]
	{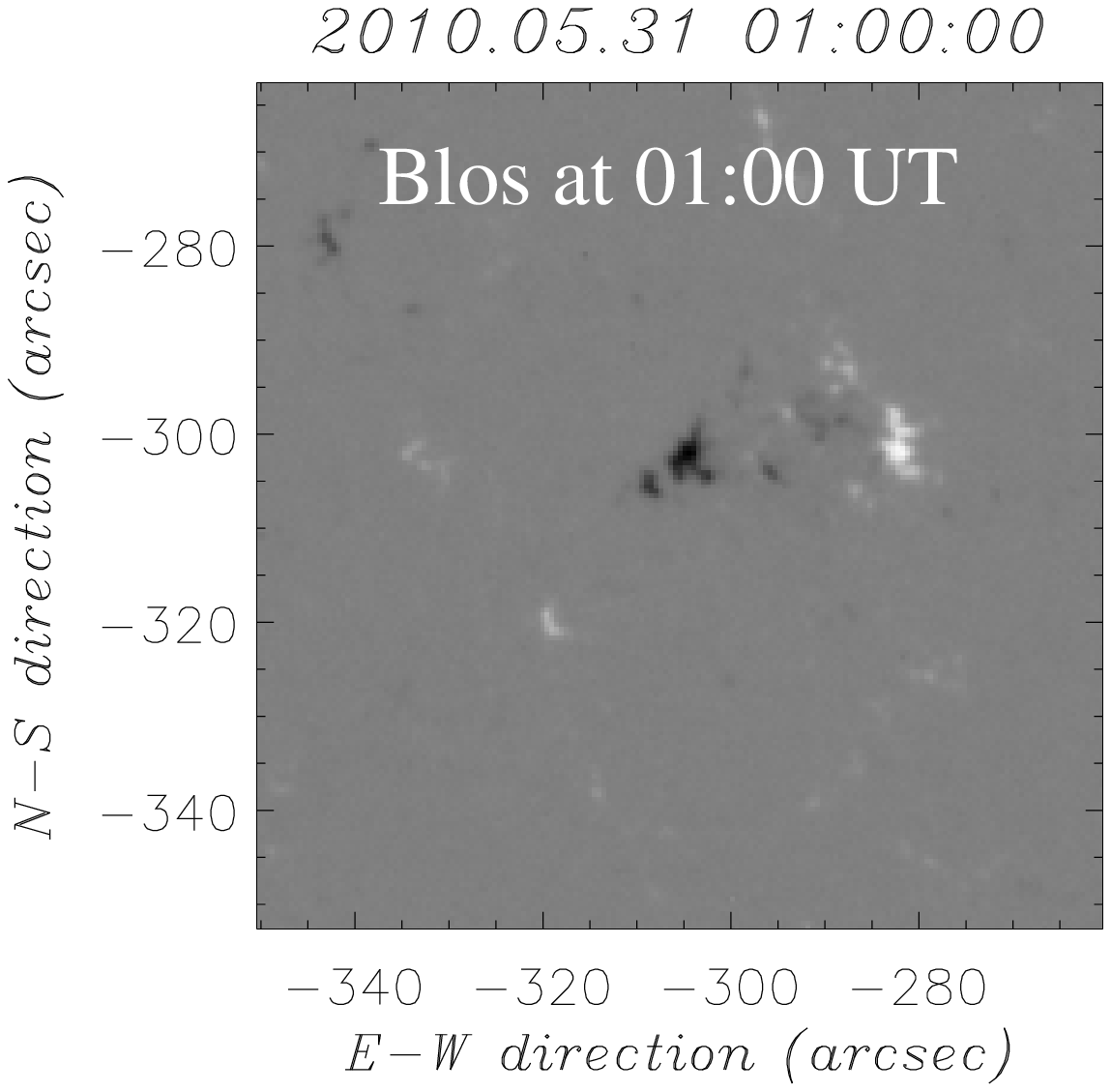}
\hfill
% 304
\includegraphics[width=.327\linewidth, bb=135 60 370 295, clip]
	{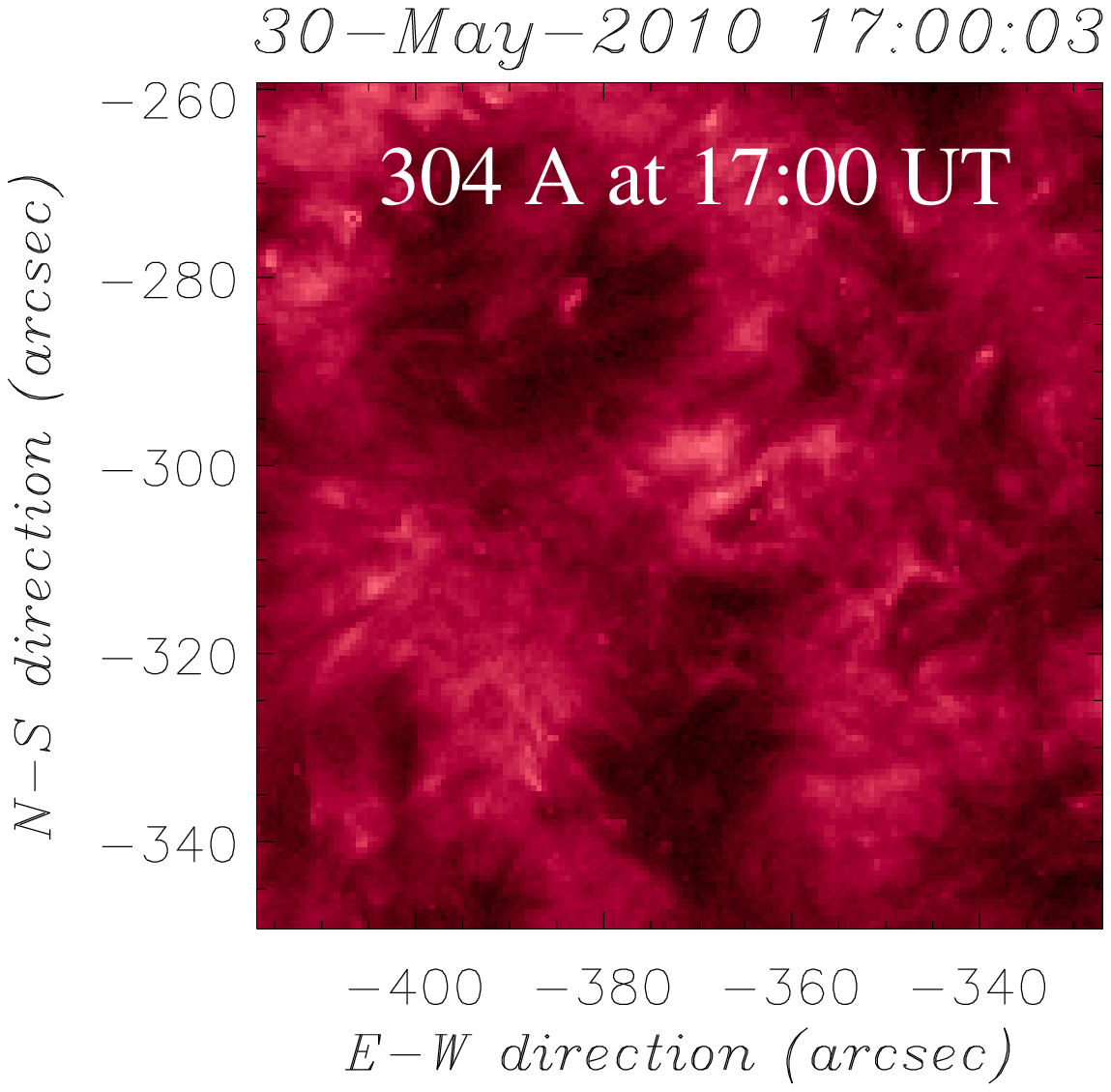}
\includegraphics[width=.327\linewidth, bb=135 60 370 295, clip]
	{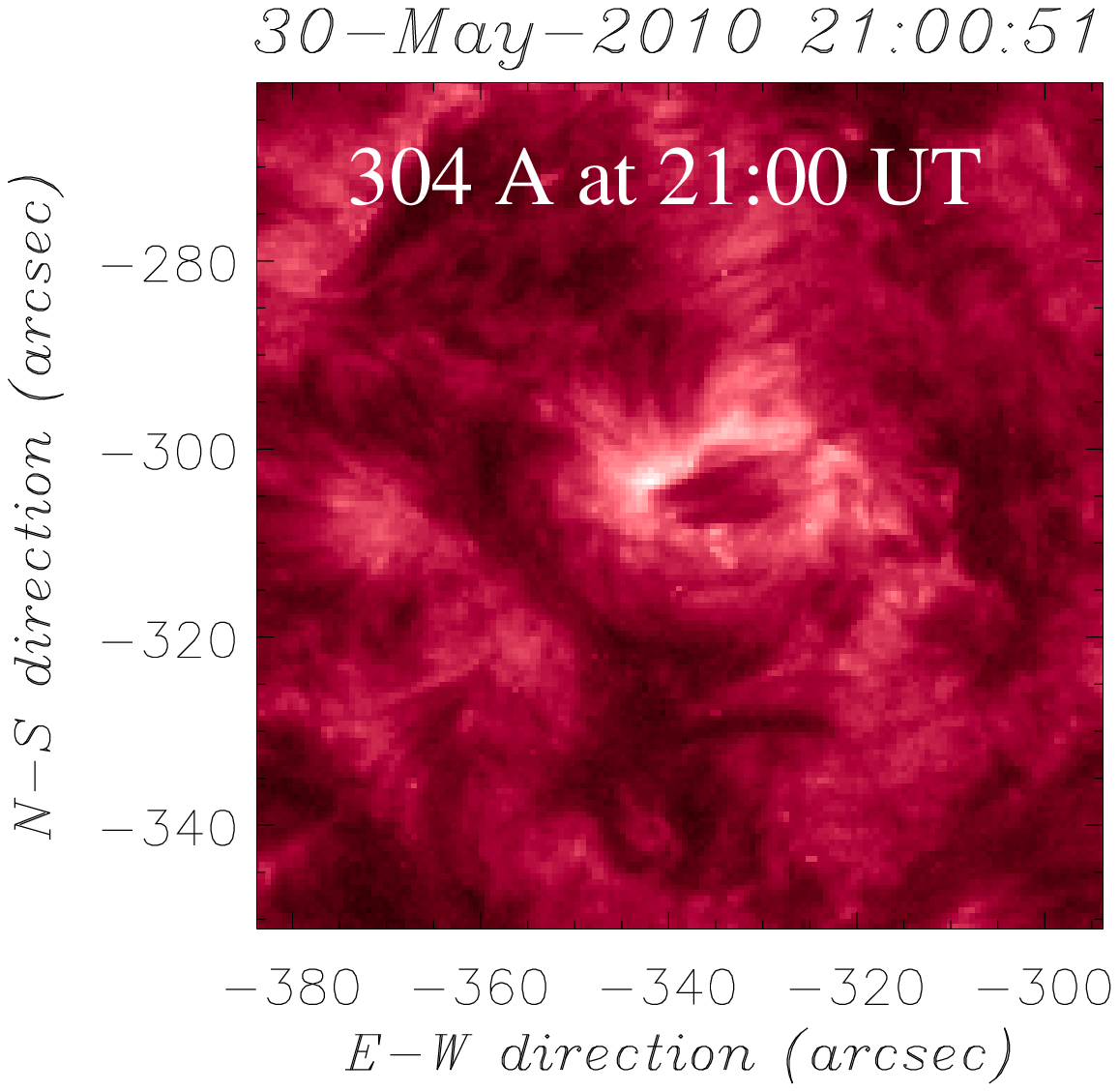}
\includegraphics[width=.327\linewidth, bb=135 60 370 295, clip]
	{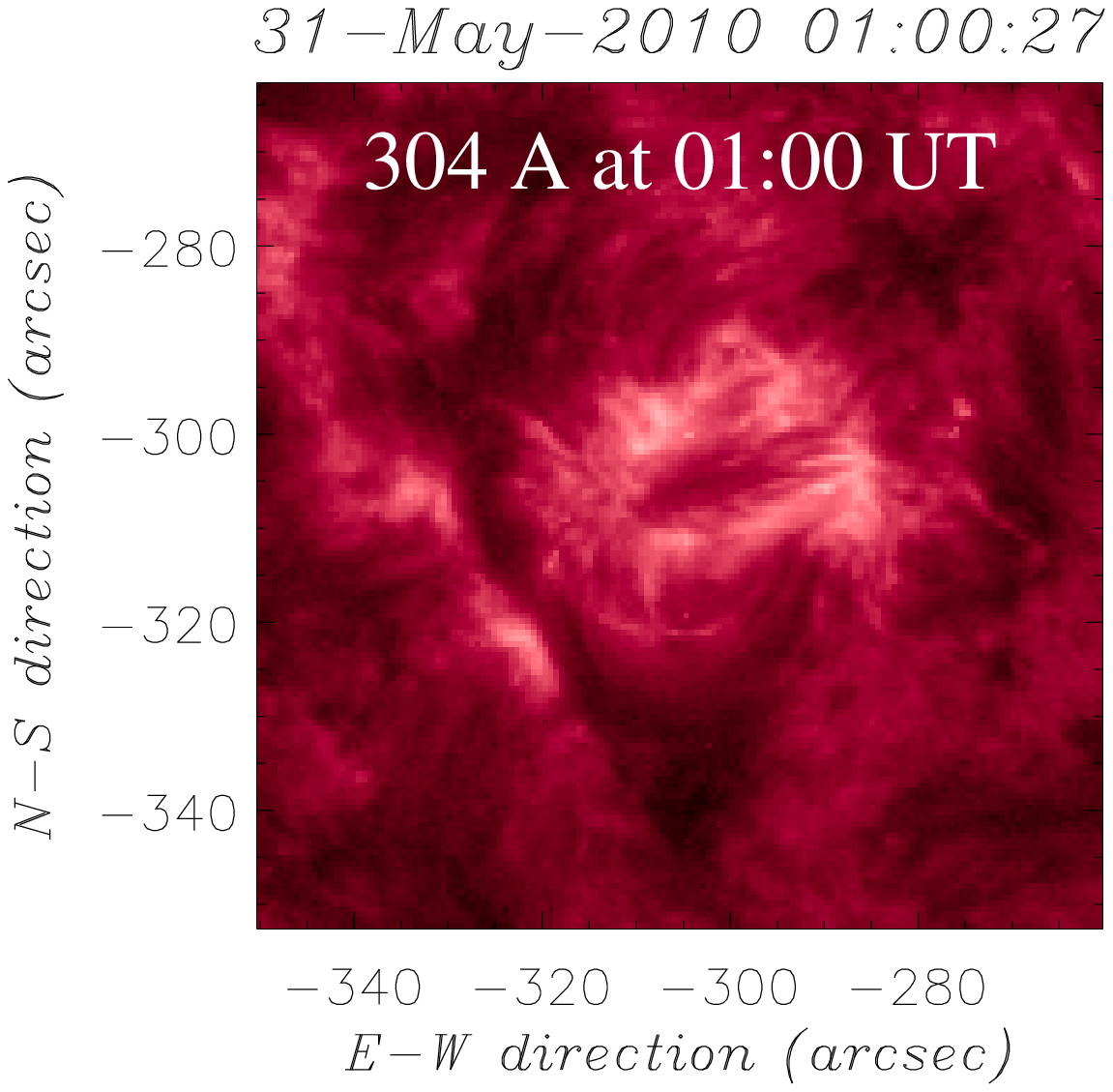}
\hfill
% 171
\includegraphics[width=.327\linewidth, bb=135 60 370 295, clip]
	{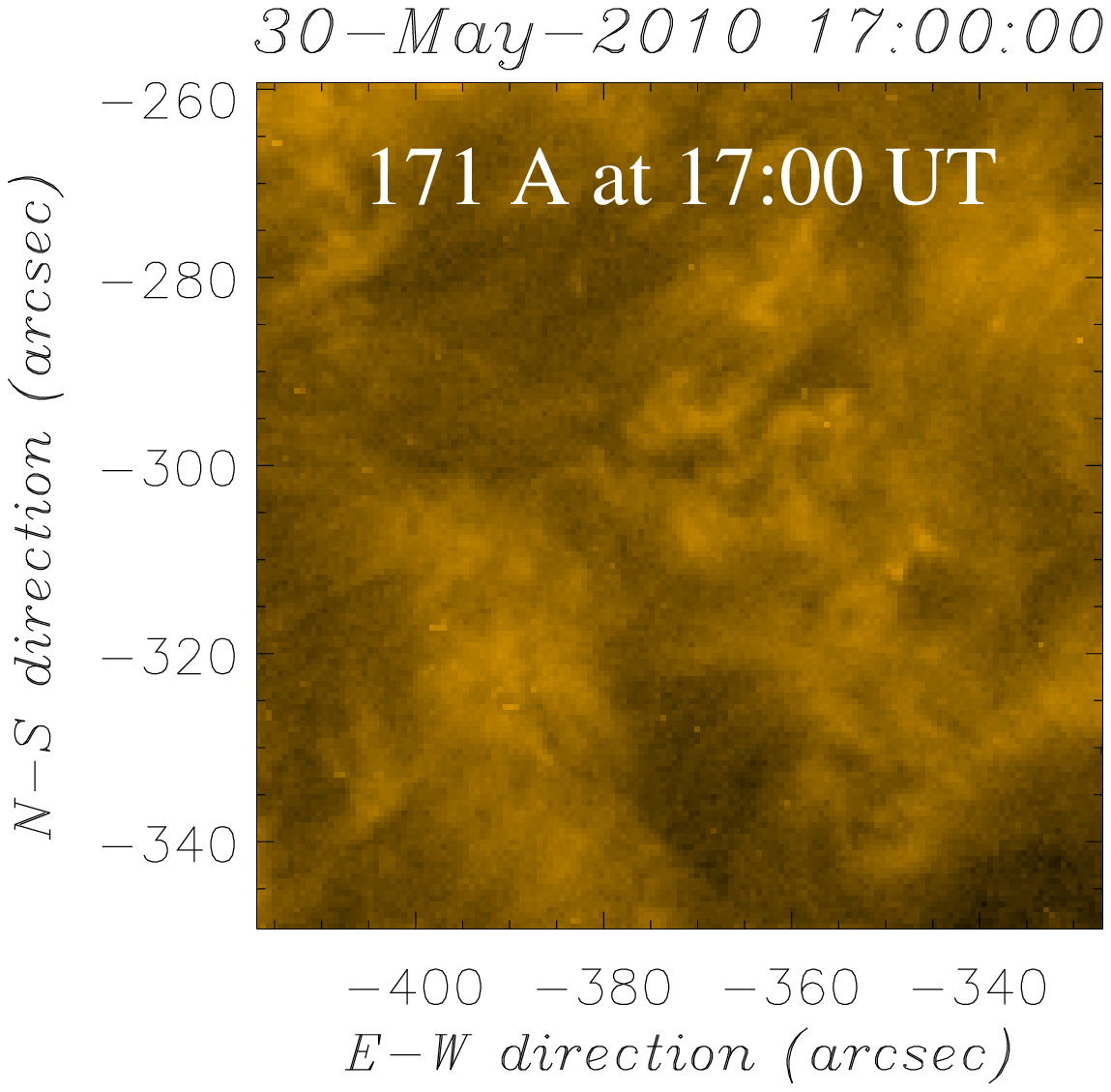}
\includegraphics[width=.327\linewidth, bb=135 60 370 295, clip]
	{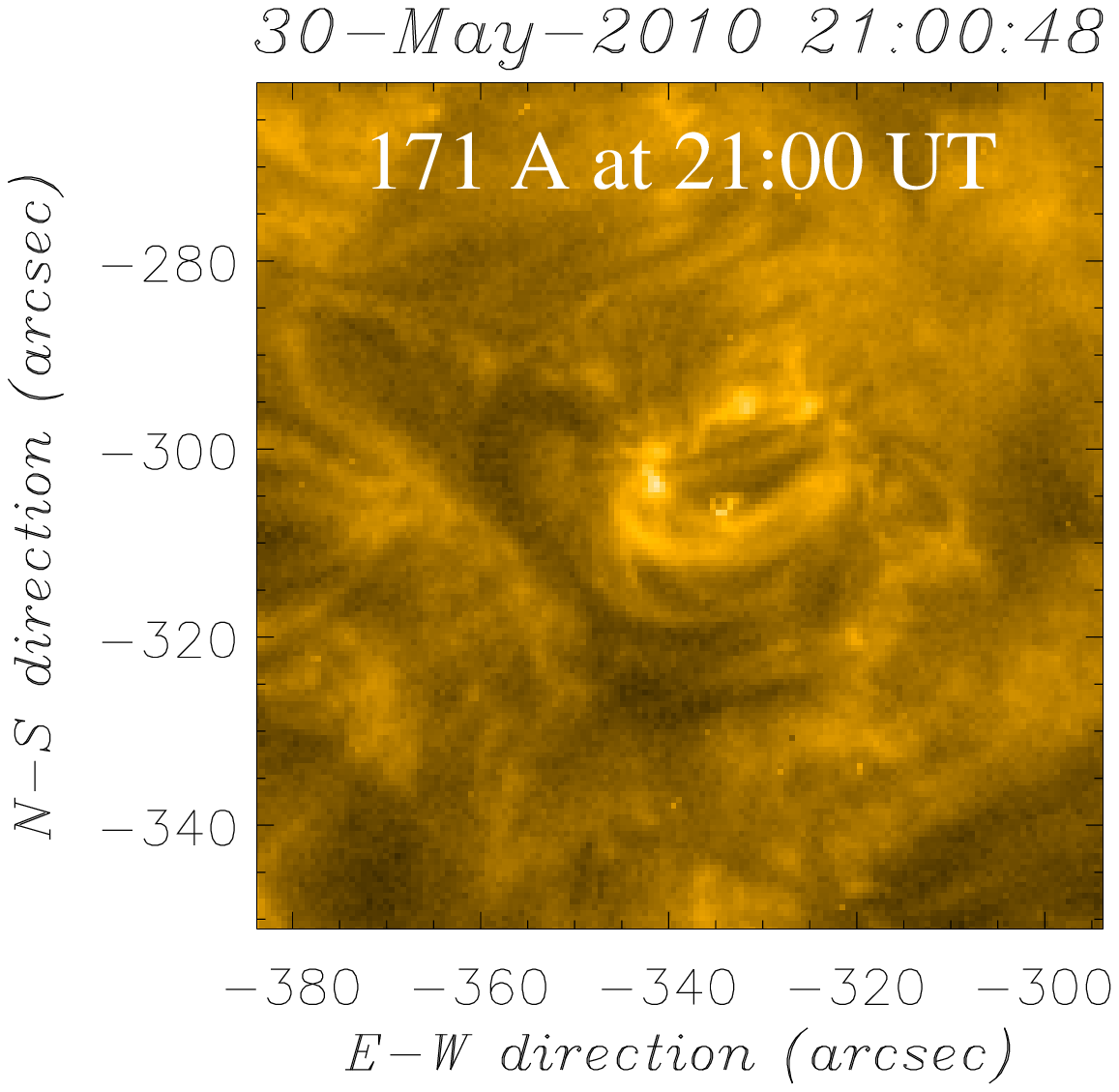}
\includegraphics[width=.327\linewidth, bb=135 60 370 295, clip]
	{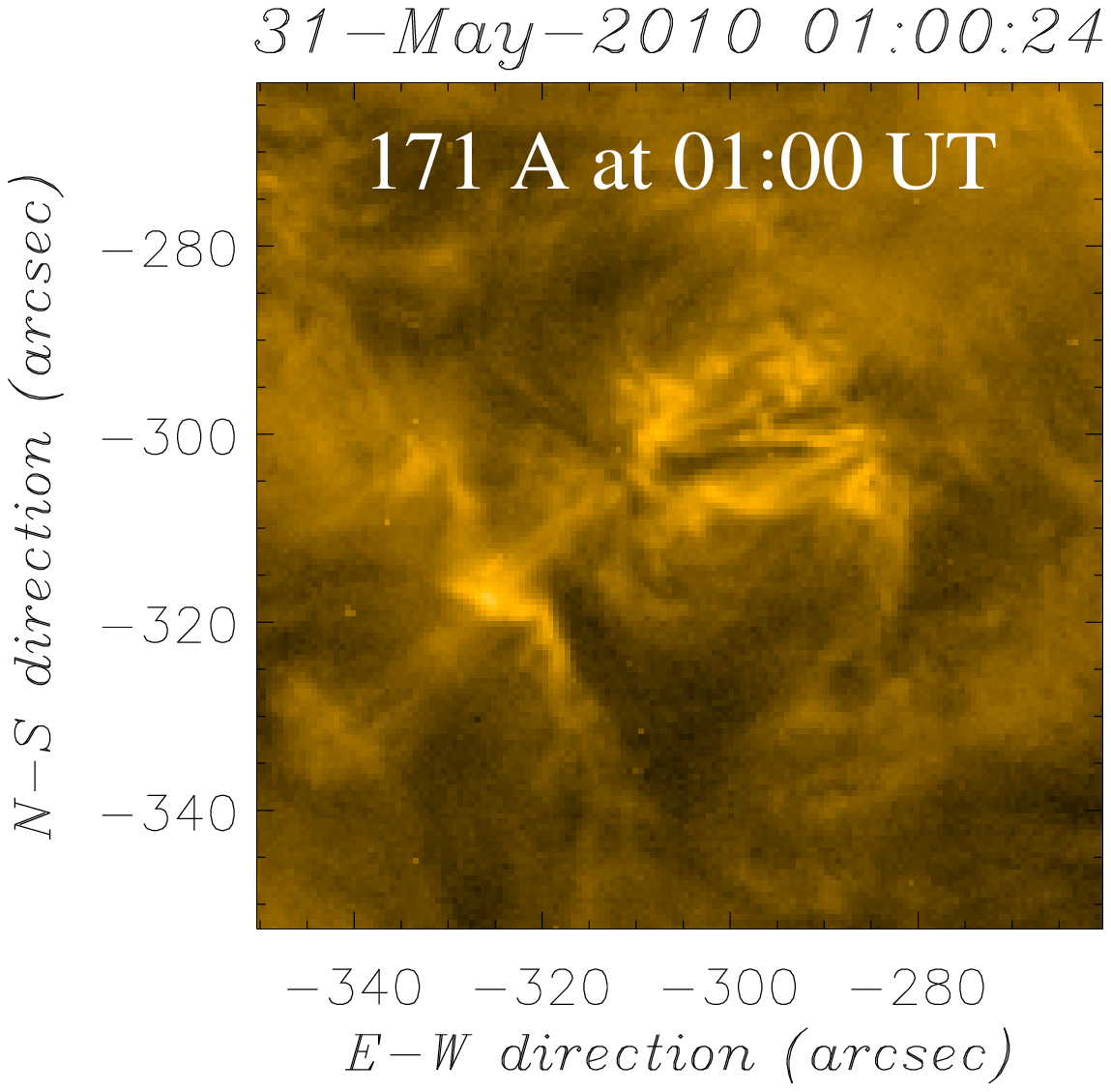}
\hfill
% 193
\includegraphics[width=.327\linewidth, bb=135 60 370 295, clip]
	{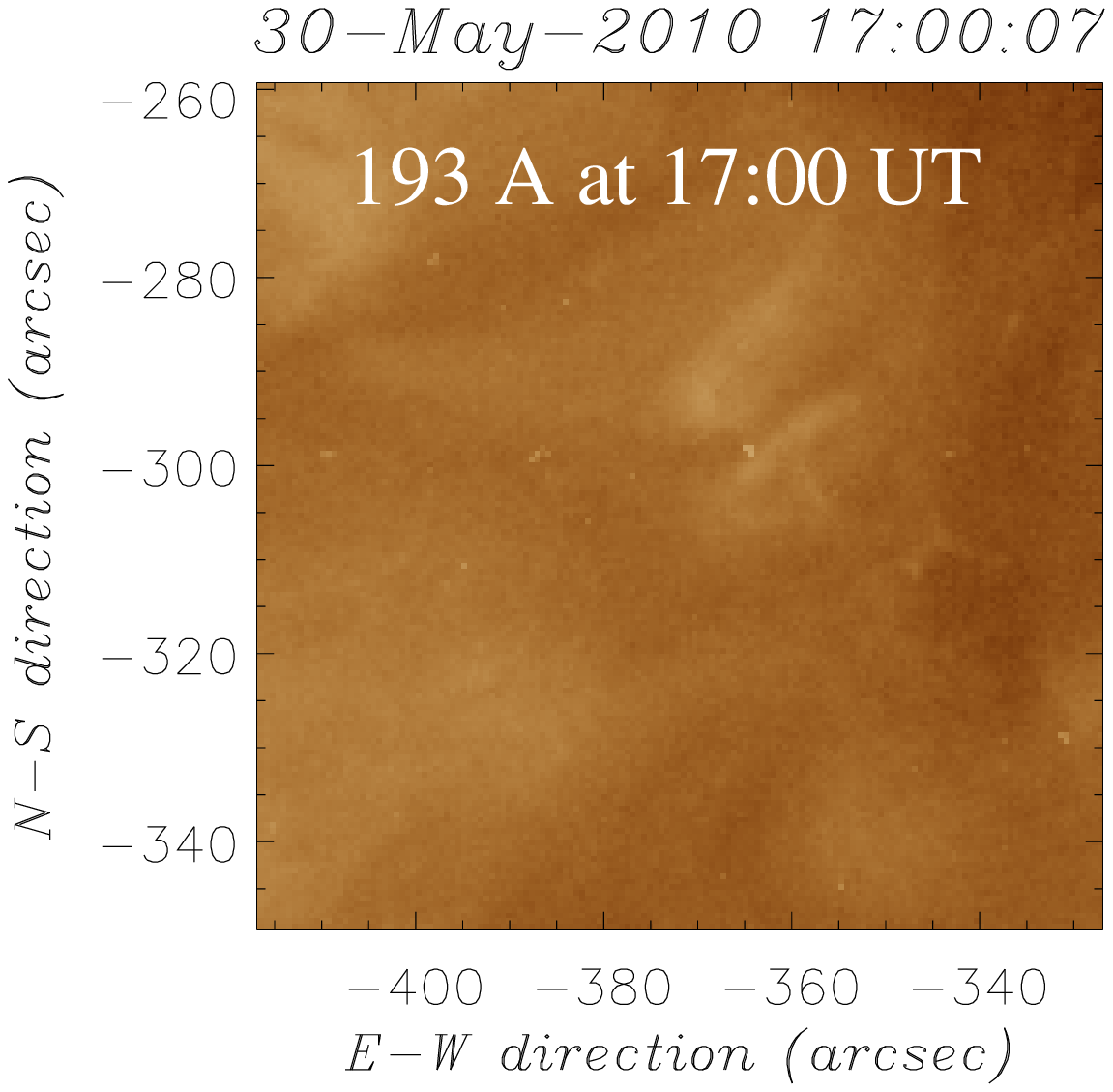}
\includegraphics[width=.327\linewidth, bb=135 60 370 295, clip]
	{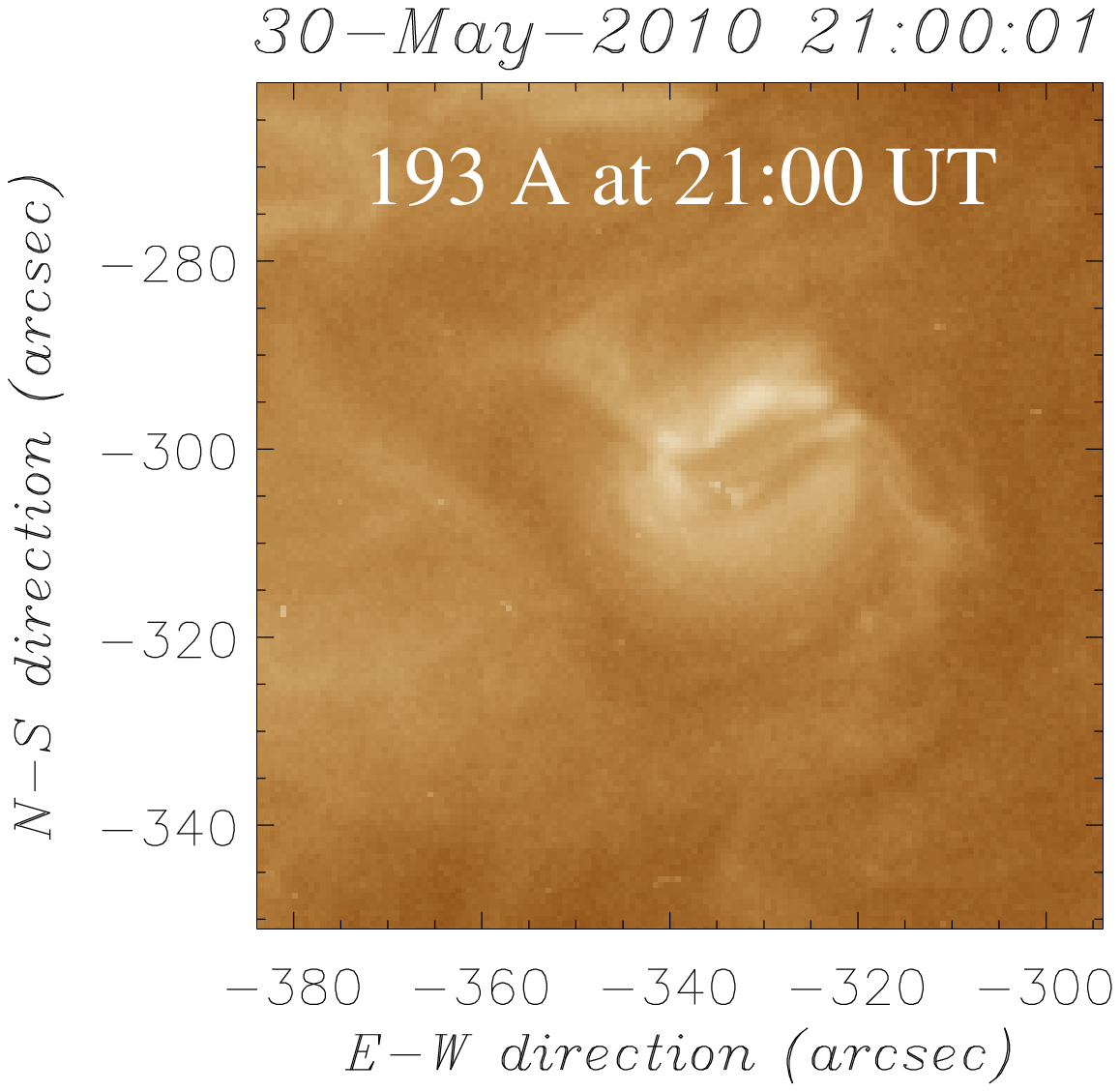}
\includegraphics[width=.327\linewidth, bb=135 60 370 295, clip]
	{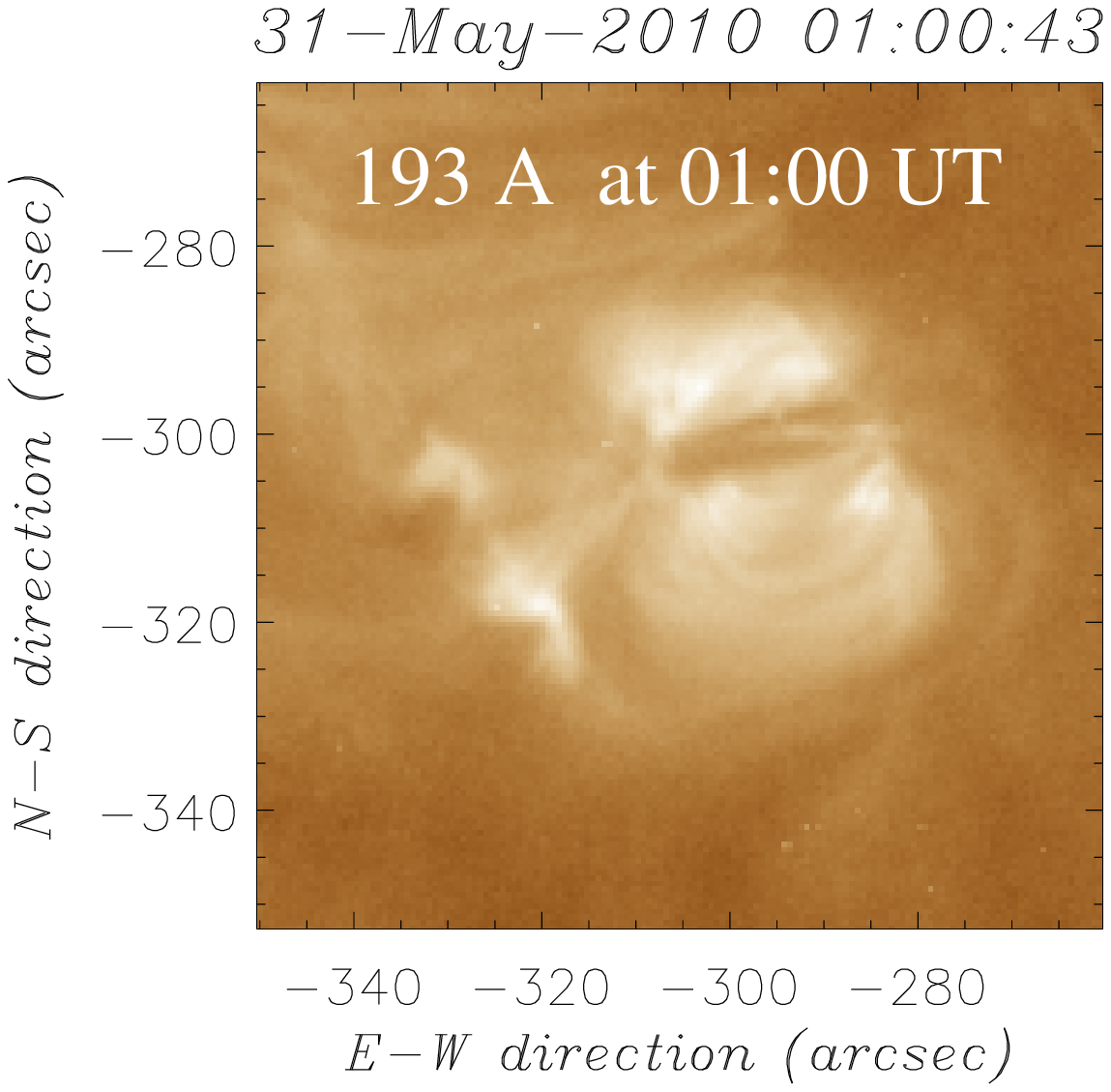}
\hfill
% 211
\includegraphics[width=.327\linewidth, bb=135 60 370 295, clip]
	{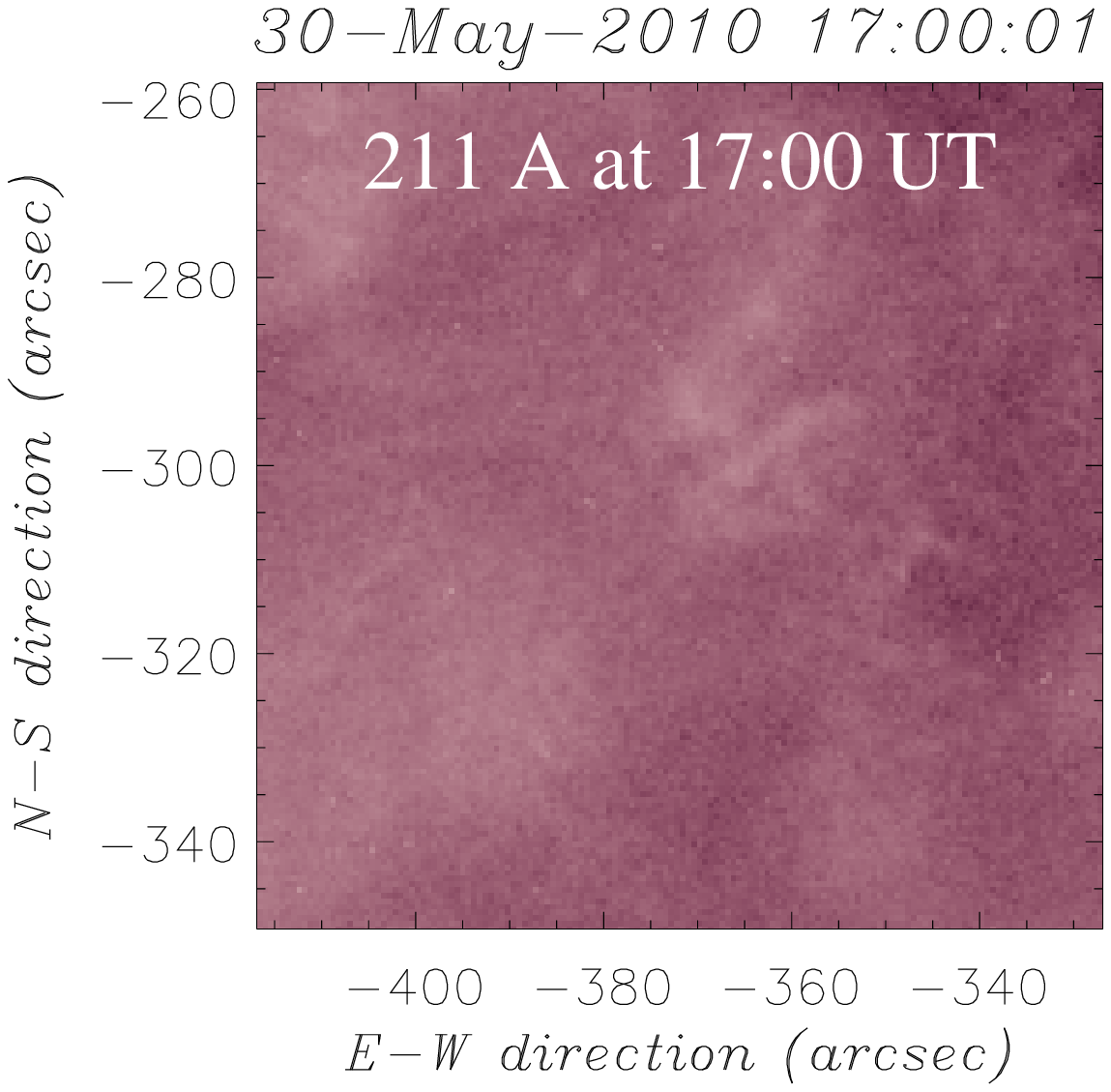}
\includegraphics[width=.327\linewidth, bb=135 60 370 295, clip]
	{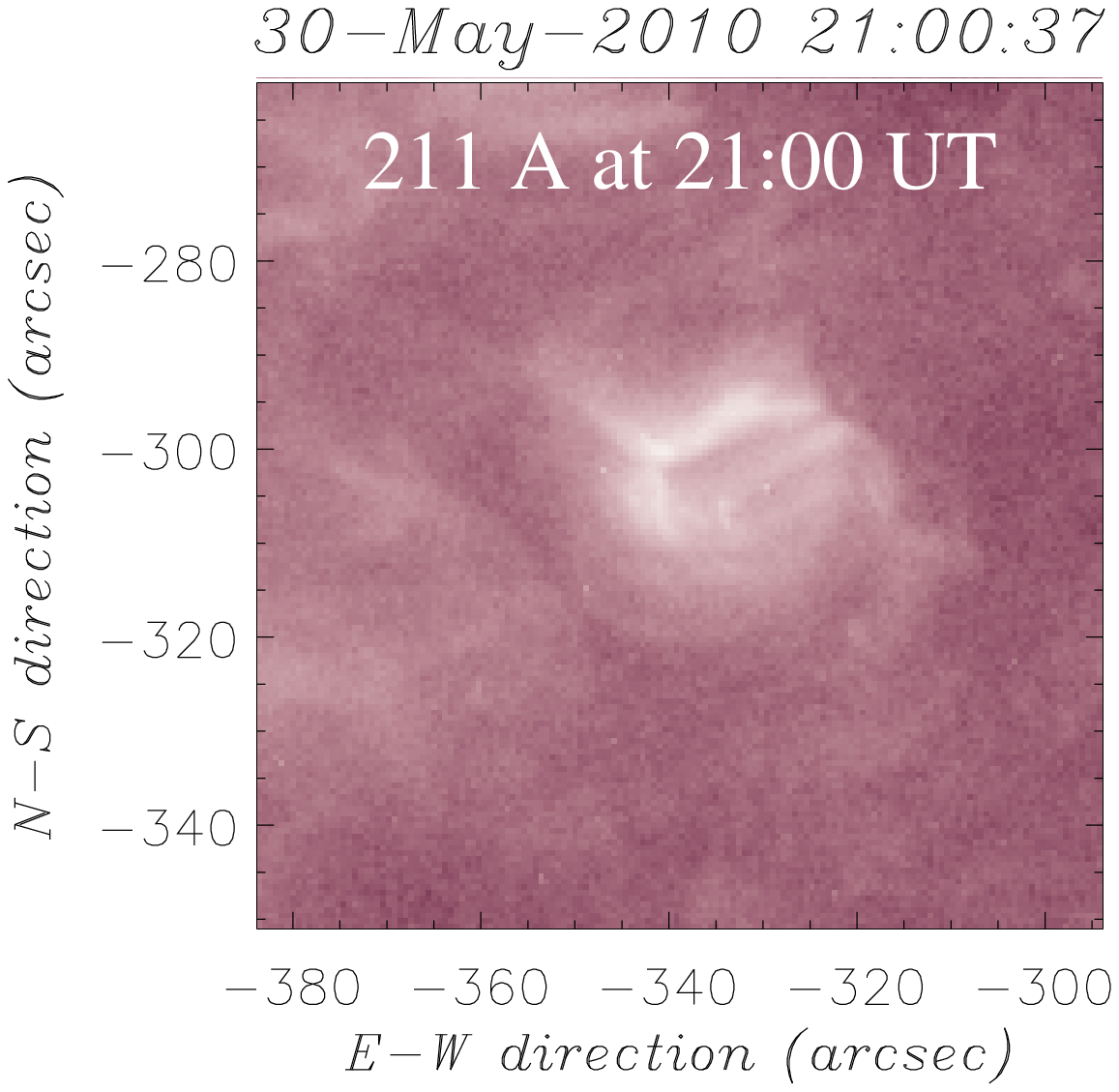}
\includegraphics[width=.327\linewidth, bb=135 60 370 295, clip]
	{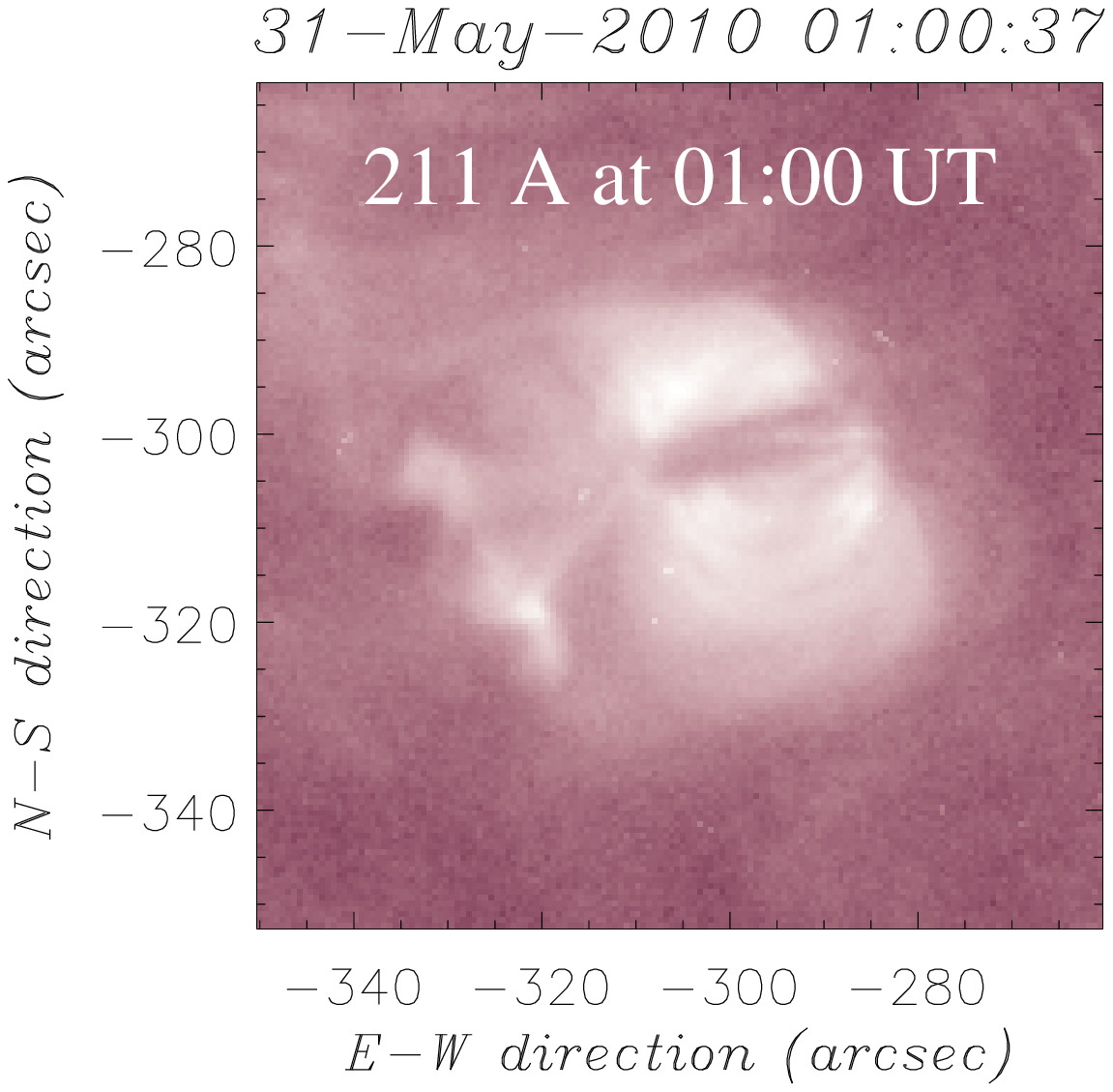}
\hfill
% 94
\includegraphics[width=.327\linewidth, bb=135 60 370 295, clip]
	{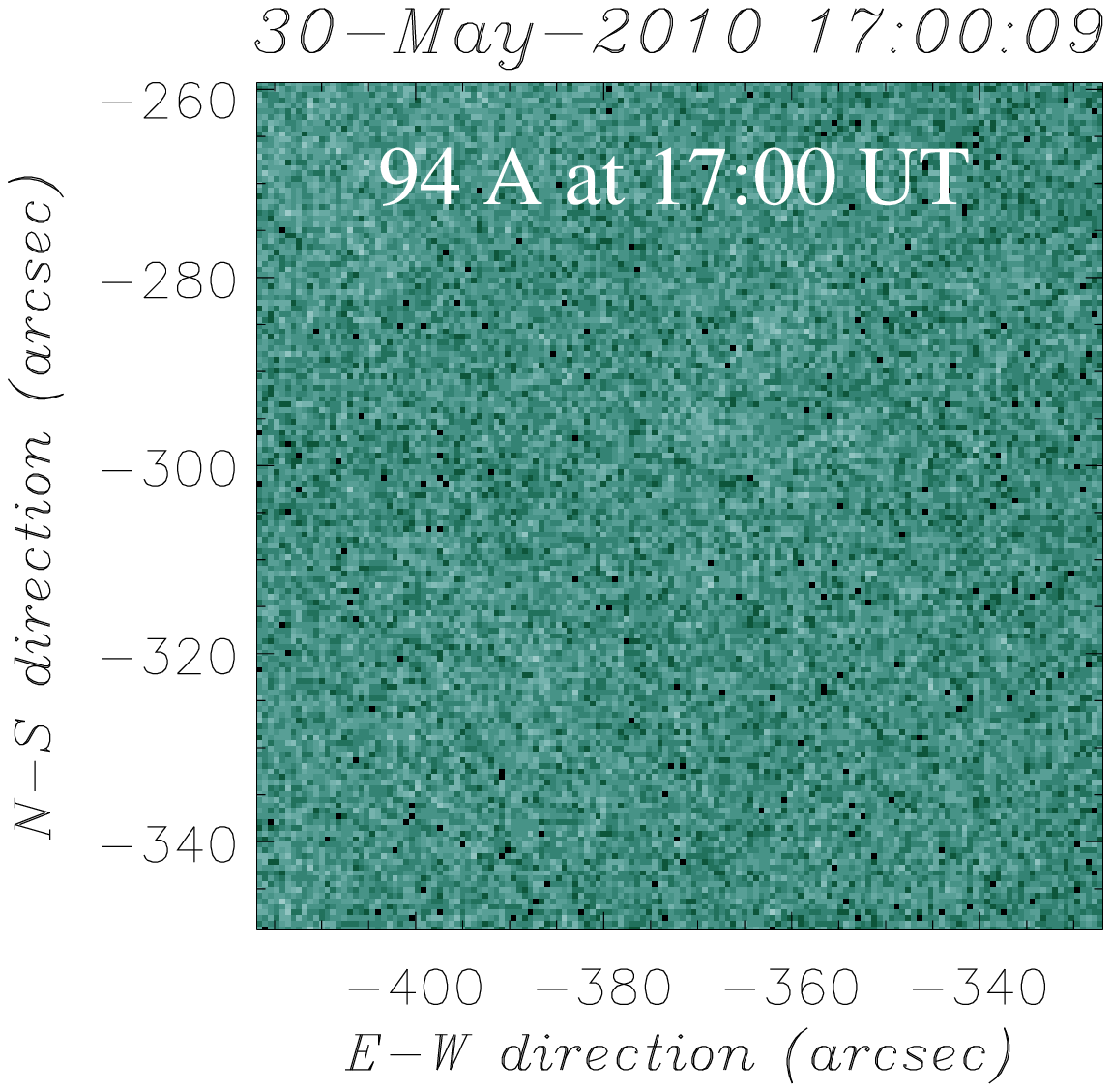}
\includegraphics[width=.327\linewidth, bb=135 60 370 295, clip]
	{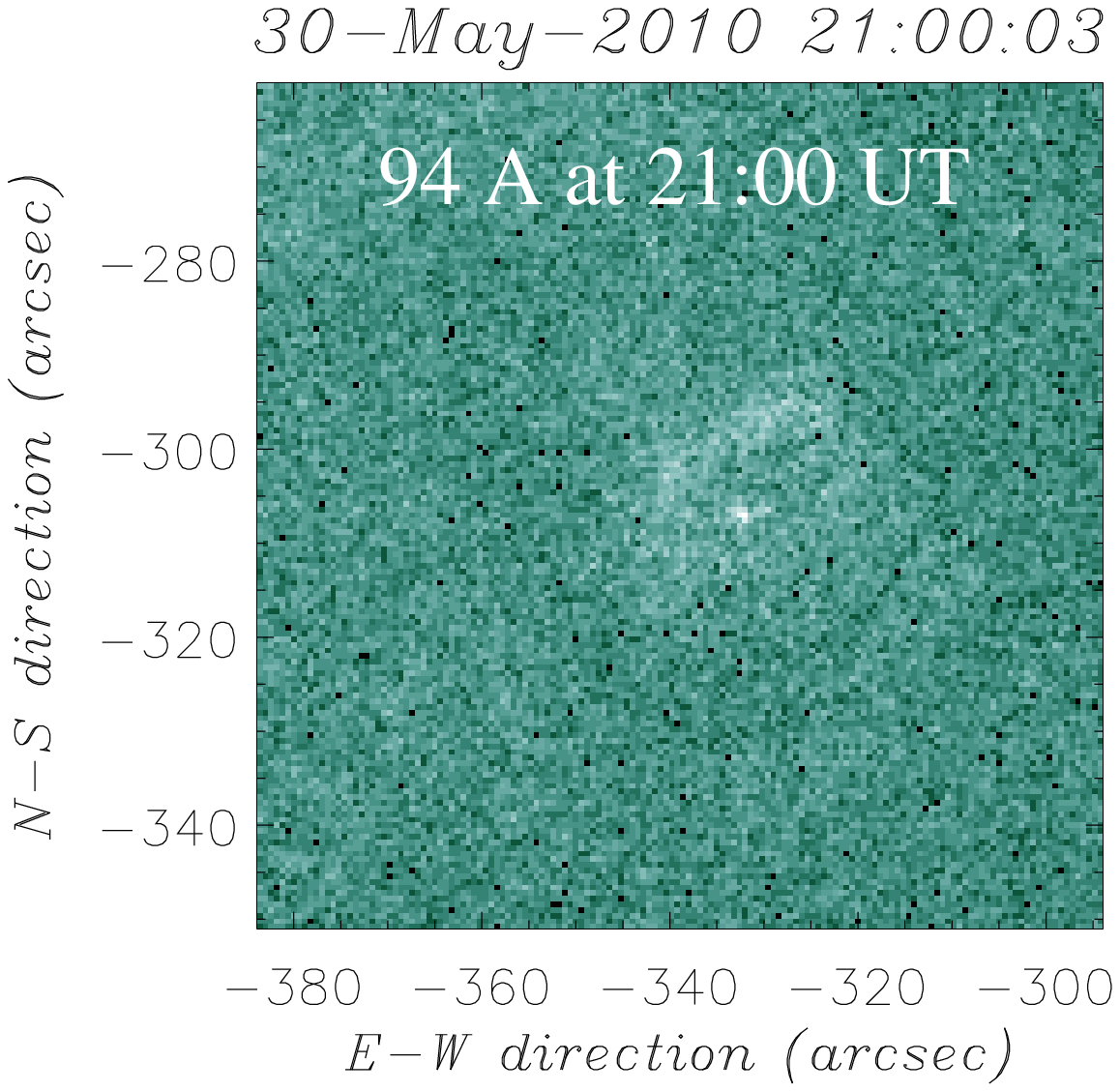}
\includegraphics[width=.327\linewidth, bb=135 60 370 295, clip]
	{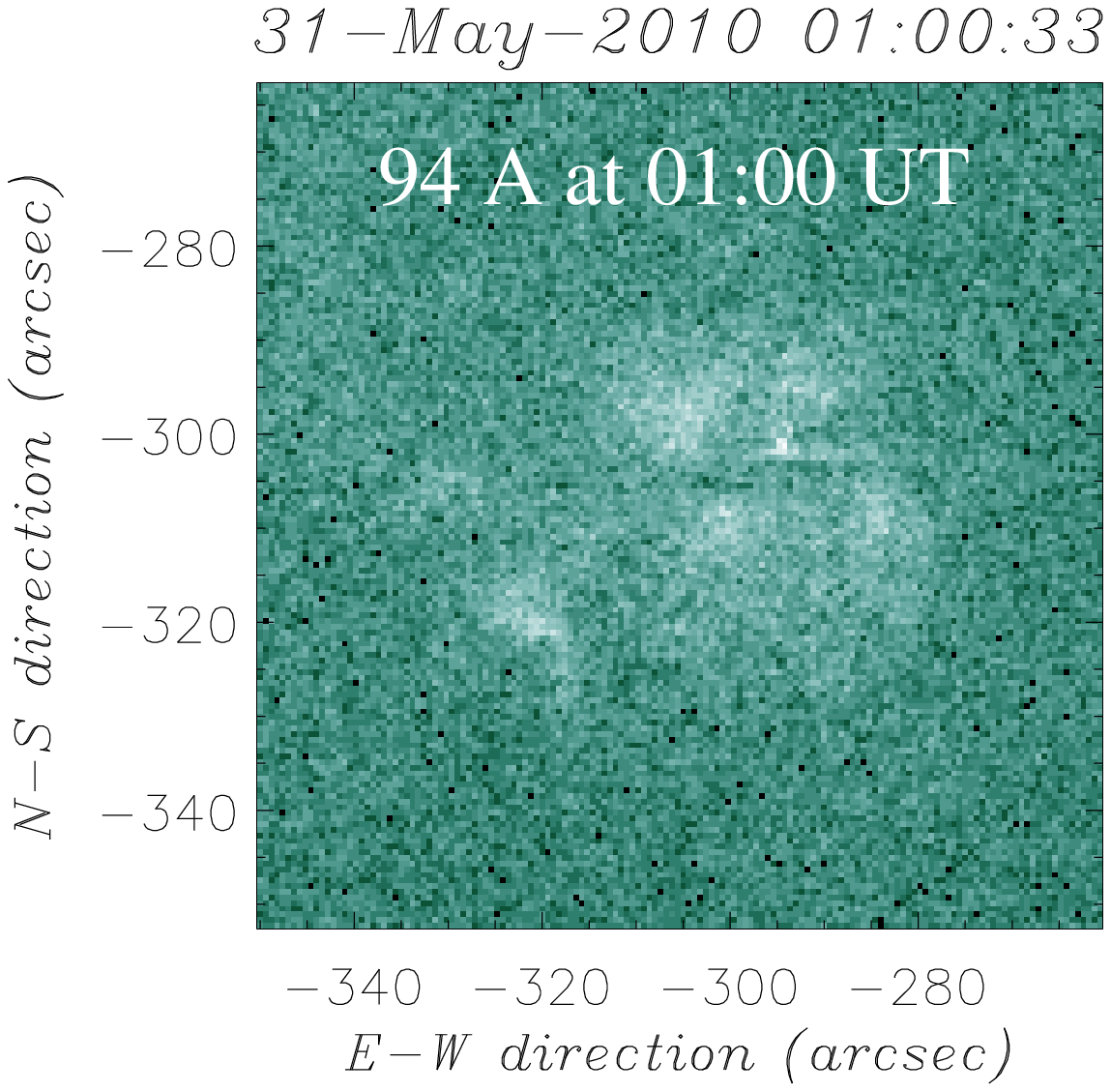}
\hfill

\caption{Time series of images showing the emergence of the active region at
three different times during the eight hours: the line-of-sight magnetic field
from SDO/HMI (top row) scaled between -800 G and 800 G, chromospheric and coronal response from SDO/AIA at
304~\AA, 171~\AA, 193~\AA, 211~\AA~and 94~\AA~(from top to bottom). The
field-of-view is 90\arcsec~in both directions centered on (-370\arcsec,
-305\arcsec) at 17:00 UT. See text for details. Movies of the time
evolution of the emergence are provided as online material.}
\label{fig:thermal}
\end{figure}

\begin{figure}
\centering
\includegraphics[width=\linewidth, bb=20 10 500 340]{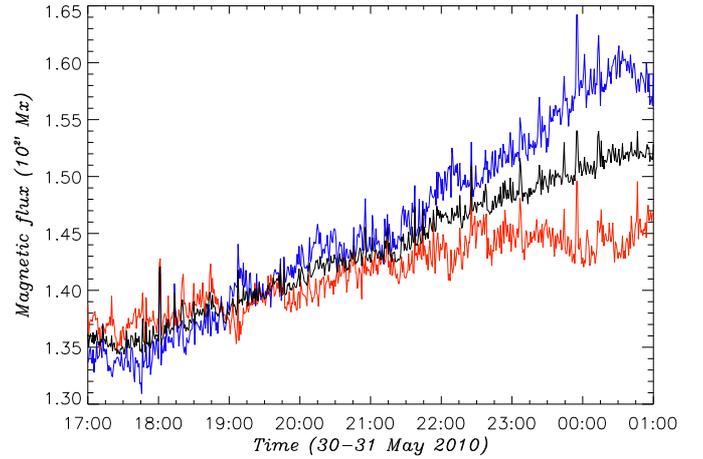}
\caption{Total magnetic flux evolution over the area of Fig.~\ref{fig:thermal}
during eight hours: the unsigned flux in black (divided by a factor of two), the
negative flux in blue (in absolute value), and the positive flux in red. The
magnetic flux is in unit of 10$^{21}$ Mx.}
\label{fig:totflux}
\end{figure}

% SDO/HMI magnetic evolution
In Fig.~\ref{fig:thermal} top row, we display the evolution of the photospheric
SDO/HMI line-of-sight magnetic field at three different times (17:00 UT, 21:00
UT, and 01:00 UT). The quiet-Sun magnetic-field distribution at 17:00 UT looks
like a supergranule magnetic field (large-scale convection cell), and
corresponds to the pre-existing coronal magnetic configuration. In
Fig.~\ref{fig:thermal} top left, the supergranular boundaries with the largest
magnetic-field strength are highlighted with ellipses. The supergranular-like
magnetic field is not unipolar but composed of mixed positive and negative
polarities. The bipolar structure of AR 11076 appears clearly in the next two
frames with increasing magnetic field strength and area. In
Fig.~\ref{fig:thermal} middle top row, we annotate the different polarities
important for the evolution: N1, P1 and P3 denote the negative (N) and positive
(P) polarities of the pre-existing quiet-Sun magnetic field, and N2 and P2 are
the polarities associated with the emerging active region. We note that, during
the first eight hours of the emergence, the direction between N2 and P2
polarities remains the same, along the east-west direction. To complement on the
evolution of the photospheric magnetic field, we plot in Fig.~\ref{fig:totflux}
the total unsigned magnetic flux (black curve), the positive (in red) and
negative (in blue) magnetic fluxes. The magnetic fluxes are computed from the
line-of-sight component of the magnetic field transformed into a vertical
component. It is noticeable that the total unsigned flux is increased as it
should for an emerging active region, however the negative flux ($+$18\%)
contributes for large part to this increase, whilst the positive flux increases
at a slower rate ($+$8\%). This gives a rate of emergence for the whole area of
4 10$^{19}$ Mx$\cdot$h$^{-1}$ with 2.7 10$^{19}$ Mx$\cdot$h$^{-1}$ for the
negative flux and 1.2 10$^{19}$ Mx$\cdot$h$^{-1}$ for the positive flux. The
emergence process is not supposed to change the net magnetic flux, however the
observed magnetic flux does not take into account the geometry or inclination of
the magnetic field (in other words the transverse components of the magnetic
field). The eight hours of this time series are long enough to emerge a
substantial amount of magnetic flux through the photosphere forming  magnetic
flux concentrations consistent with recent numerical simulations of emerging
active regions by \citet{che10}. However, according to these authors, pores and
sunspots are forming latter during the emergence phase. These observations are
supporting the results obtained by \citet{che10}.

%%%%%%%%%%%%%%%%%%%%%%%%%%%%%%%%%%%%%%%%%%%%%%%%%%%%%%%%
%%%%%%%%	Thermal Structure Images	%%%%%%%%
%%%%%%%%%%%%%%%%%%%%%%%%%%%%%%%%%%%%%%%%%%%%%%%%%%%%%%%%

%%%%%%%%%%%%%%%%%%%%%%%%%%%%%%%%%%%%%%%%%%%%%%%%%%%%%%%%

%%%%%%%%%%%%%%%%%%%%%%%%%%%%%%%%%%%%%%%%%%%%%%%%%%%%%%%%
%%%%%%%%	Magnetic flux evolution		%%%%%%%%
%%%%%%%%%%%%%%%%%%%%%%%%%%%%%%%%%%%%%%%%%%%%%%%%%%%%%%%%

%%%%%%%%%%%%%%%%%%%%%%%%%%%%%%%%%%%%%%%%%%%%%%%%%%%%%%%%

% SDO/AIA thermal shielding
In Fig.~\ref{fig:thermal} rows 2-6, the EUV emission lines show the behaviour of
the emerging active region from 50000K to several million degrees. The intensity
maps of the 304\AA~and 171\AA~channels show a lot of fine structuring within the
emerging flux including dark material which is presumably cool material. The
emission for the other three channels is more diffuse (and near the noise level
for the 94\AA~channel). From the time series (Fig.~\ref{fig:thermal}), it is
clear that hot channels such as 193\AA~, 211\AA, and 94\AA~show the large-scale
magnetic field lines of the emerging flux bundles and thus interfacing more with
the pre-existing magnetic field. We determine the expansion of the emerging
flux in the north-south and east-west directions by measuring the full-width at
half-maximum of the intensity variation at a given time. As expected, the
expansion of the emerged magnetic field is larger in regions of weak magnetic
field strength (north-south direction in this case) than along the main axis of
the active region (east-west direction): 
\begin{itemize}
\item[-]the length in the north-south
direction is about 18 Mm at 21:00 UT and 28 Mm at 01:00 UT. The rate of
expansion in the north-south direction can then be estimated at 2.5
Mm$\cdot$h$^{-1}$. This rate of expansion will give a typical active region of
size 120 Mm after two days; 
\item[-]there is no measurable expansion in the east-west direction.
\end{itemize}

EUV brightenings in the south-east part of the
active region (coordinates X=-320\arcsec, Y=-320\arcsec in
Fig.~\ref{fig:thermal} rows 4 and 5 at 01:00 UT) show the interaction between
the pre-existing coronal field and the emerging flux creating magnetic
connection between the two polarities (P3 and N2) as evidenced by the EUV loops. 

%%%%%%%%%%%%%%%%%%%%%%%%%%%%%%%%%%%%%%%%%%%%%%%%%%%%%%%%
%%%%%%%%	Thermal Shielding lightcurve	%%%%%%%%
%%%%%%%%%%%%%%%%%%%%%%%%%%%%%%%%%%%%%%%%%%%%%%%%%%%%%%%%

\begin{figure}
\centering
\includegraphics[width=1.\linewidth, bb=35 10 480 360]{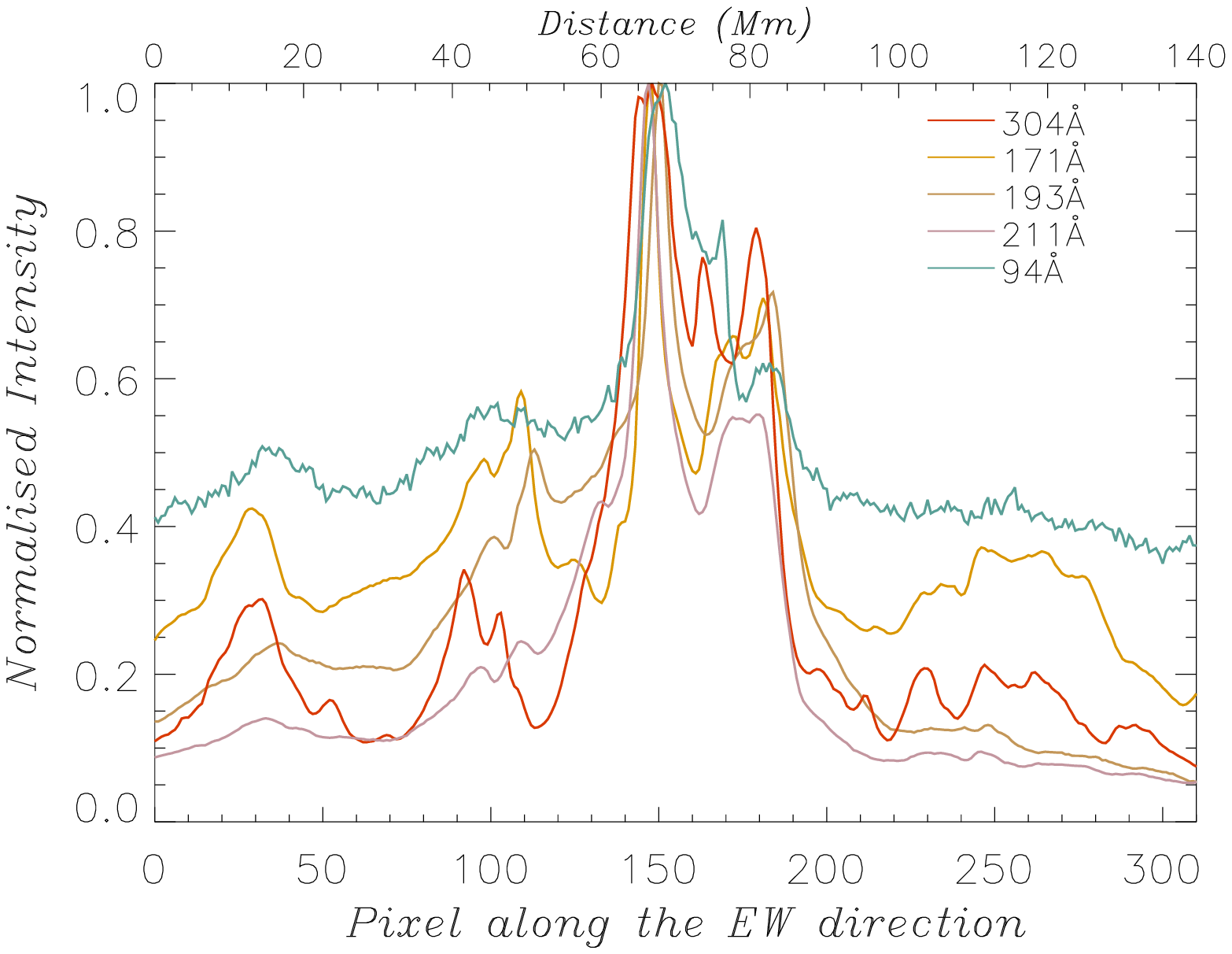}
\includegraphics[width=1.\linewidth, bb=35 10 480 360]{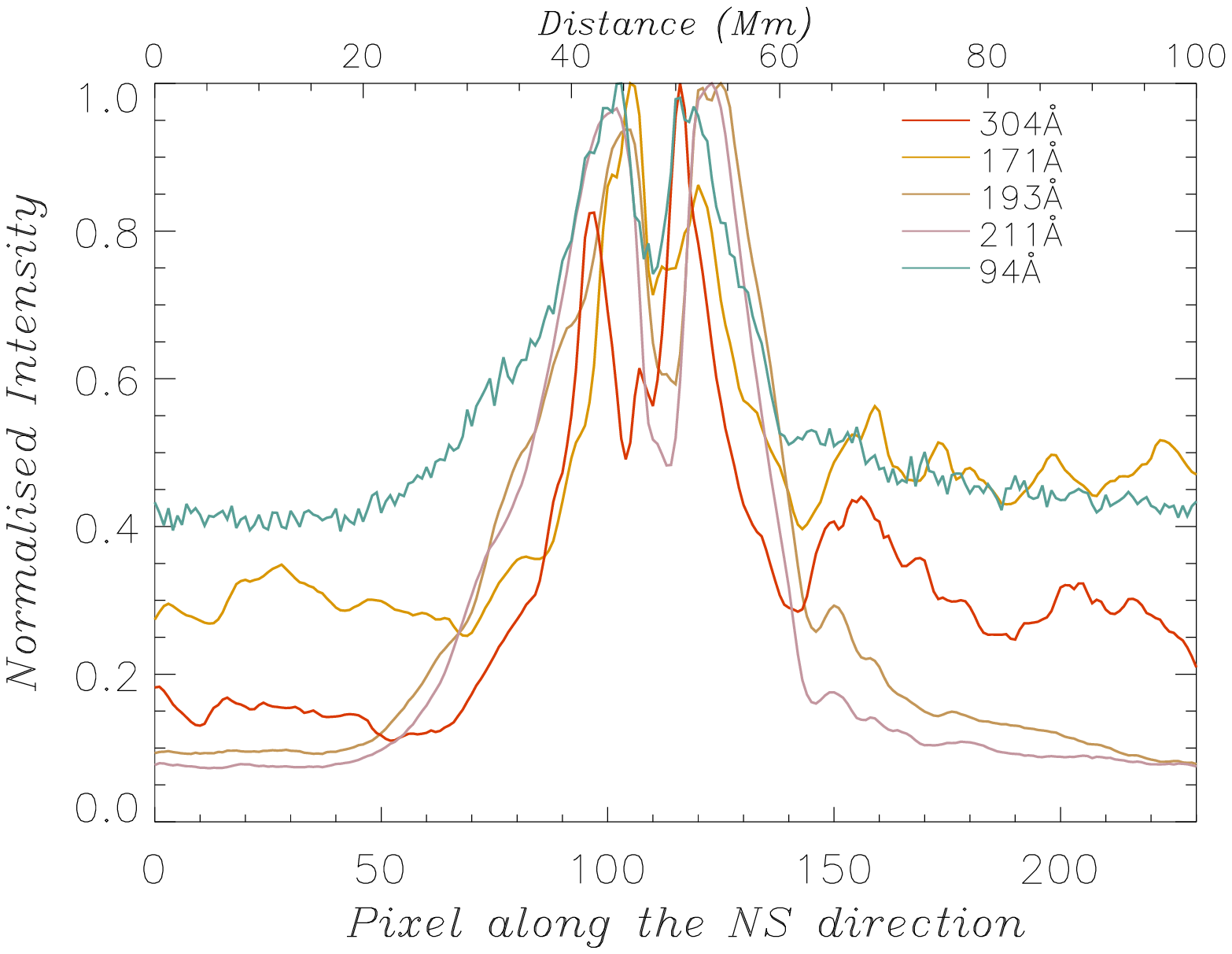}
\caption{Intensity variation averaged over three hours (22:00 to 01:00 UT) for
all five SDO/AIA channels: ({\em top}) along the east-west direction at the NS
location of -300\arcsec,  and ({\em bottom}) along the south-north direction at
the initial EW location of -375\arcsec. The intensity is normalised to the
maximum intensity for the sake of comparison. }
\label{fig:lc}
\end{figure}

%%%%%%%%%%%%%%%%%%%%%%%%%%%%%%%%%%%%%%%%%%%%%%%%%%%%%%%%
To investigate the location and process of the heating, we plot the intensity
variation for five SDO/AIA channels averaged over three hours (from 22:00 to
01:00 UT). In Fig.~\ref{fig:lc} top, the intensity variations in the east-west
direction along the axis between the negative and positive polarities of the
emerging region are plotted. All five curves have basically the same shape with
a clear double peak structure where the edges of the emerging flux region are
located. The peaks are not symmetric: the peak near the negative polarity
(left-hand side) is at least 20\% higher than the peak associated with the
positive polarity (right-hand side). This asymmetry is important to determine
the processes responsible for heating the emerging configuration. As the
magnetic field strength is similar in both polarities (about 800 G), the
asymmetry indicates two different processes: (i) magnetic reconnection
converting the magnetic energy into kinetic and thermal energy with magnetic
field lines having different orientations, and (ii) magnetic compression due to
the increase of magnetic pressure when the geometry of the field does not allow
magnetic reconnection to occur. In Fig.~\ref{fig:bemerg}, we draw a 2D sketch
summarising the processes at play during the emergence of the active region and
producing the asymmetry in the double peak emission. As indicated in
Fig.~\ref{fig:thermal} top row, N2 and P2 form the emerging bipolar region
corresponding to the field lines in red. We notice that, on the right-hand side,
the field lines are anchored in the positive polarities (P1 and P2) with the
same orientation, which is more favorable for an heating mechanism such as the
magnetic compression. On the left-hand side of the emerging flux, the magnetic
topology is more complex and it is more likely to produce magnetic reconnection.
The highest peak being on the left-hand side is explained by the impulsive
nature of the magnetic reconnection compared to the slow process of magnetic
compression. In the movie showing the evolution in the 171\AA\ channel, we
observe that magnetic field lines originating in N2 are reconnected from P2 to
P3. 

In Fig.~\ref{fig:lc} bottom, we plot the intensity variation in the
south-north direction perpendicular to the axis of the active region. We find
again the double peak structure, however the intensity variations are symmetric:
this is more likely to indicate a heating mechanism such as magnetic
compression. A common feature to both intensity variation plots is the double
peak structure: two peaks on the side which can be explained by either magnetic
reconnection and magnetic compression, and an emission dip. The dip is caused by
the emergence of cool plasma appearing as dark material in the EUV images of
Fig.~\ref{fig:thermal}. This is also a common feature in models of magnetic flux
emergence \citep{mag01,arc07,mar08} in which the emerging flux rope has cool
material at its centre. These observations of a newly
emerged flux show similar time-scale and spatial-scale than the simulation of
\cite{che10}, only the magnetic field strength is noticeably different (800 G in
the observation, 3000 G in the simulation).        

%%%%%%%%%%%%%%%%%%%%%%%%%%%%%%%%%%%%%%%%%%%%%%%%%%%%%%%%%%%%%%%%%%%%%%%%%%%
\section{Discussion and Conclusions}
\label{sec:disc}
%%%%%%%%%%%%%%%%%%%%%%%%%%%%%%%%%%%%%%%%%%%%%%%%%%%%%%%%%%%%%%%%%%%%%%%%%%%

We observe the first eight hours of the emergence of an active region. The
peculiarities of this flux emergence are (i) the new polarities emerge near a
supergranular-like boundary, (ii) no rotation of the polarities forming the
emerging active region with respect to each other is observed. This strongly
suggest that the emergence process is either due to the buoyancy of a flux tube
containing a small amount of twist, or associated with a flux sheet near the
solar surface.

\begin{figure}
\centering
\includegraphics[width=\linewidth]{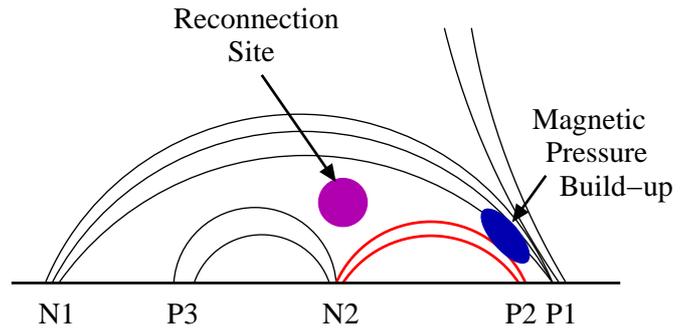}
\caption{Sketch of the magnetic field structure of the emerging active region as
deduced from the distribution of polarities in Fig.~\ref{fig:thermal} middle top
row. The field lines in red characterise the emerging magnetic field. East is on
the left-hand side.}
\label{fig:bemerg}
\end{figure}

The SDO/AIA observations provide an unprecedented view of the thermal structure
of the emerging active region and its interaction with the hot corona. In order
to analyse the thermal structure of the emerging active region, we plot the
average intensity variation for the different wavelengths for the last 3 hours of
the time series when the size of the active region remains almost constant. We
observe the thermal shielding of the emerging flux at the interface with the
pre-existing quiet-Sun magnetic field: all wavelength channels exhibit two
strong peaks at the interface between the emerging region and the pre-existing
corona. By comparing with 3D models \citep{mag01,arc07,mar08}, we deduce that
the observed asymmetry in the intensity variation between the east and west sides is
due to two different processes as depicted in Fig.~\ref{fig:bemerg}: 
\begin{itemize}
\item[(i)]{on the west side, the build-up of magnetic pressure in a system
of magnetic field lines with the same orientation is responsible for the
increase of temperature at the shield;}  
\item[(ii)]{on the east side, magnetic
reconnection can be invoke to heat the shield owing to the opposite orientation
of field lines and the complex geometry of the field.}
\end{itemize} 

We have been able to characterised the thermal structure of an emerging active
region in the quiet Sun. In addition, the magnetic flux emergence process also
generates impulsive events such as external magnetic reconnection between the
emerging and pre-existing magnetic fields, and internal magnetic reconnection
within the emerging flux bundles. These processes and how they contribute to the
emergence will be discussed in a forthcoming paper. 

%%%%%%%%%%%%%%%%%%%%%%%%%%%%%%%%%%%%%%%%%%%%%%%%%%%%%%%%
%%%%%%%%	Magnetic field schema	%%%%%%%%
%%%%%%%%%%%%%%%%%%%%%%%%%%%%%%%%%%%%%%%%%%%%%%%%%%%%%%%%

%%%%%%%%%%%%%%%%%%%%%%%%%%%%%%%%%%%%%%%%%%%%%%%%%%%%%%%%

\begin{acknowledgements}
I thank the referee for her/his comments which have improved the
manuscript. I would like to thank Vasilis Archontis for fruitful discussions on
flux emergence. The SDO data have been collected from the University of Central
Lancashire database. The data used are provided courtesy of NASA/SDO and the AIA
and HMI science teams.

\end{acknowledgements}

%clearpage

\bibliographystyle{aa}
%\bibliography{/wisdom/work/sr1/TEX/Bib/mybib}

\Online

\begin{appendix} %First online appendix
\section{Time evolution of the emerging active region}

A movie of the evolution of the emerging active region during eight hours (17:00
UT on 30 May 2010 to 01:00 UT on 31 May 2010, 36s time cadence) is supplied as
online material.

\begin{figure}[!h]
\centering
\includegraphics[width=1.\linewidth]{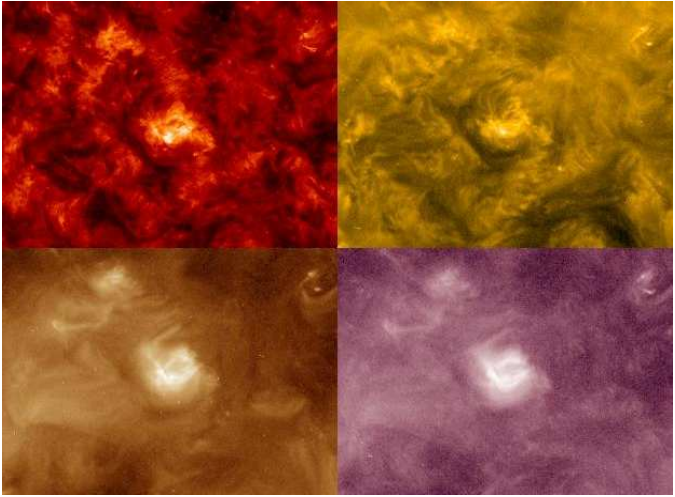}
\caption{Still image of the SDO/AIA movie at 21:30 UT on 30 May 2010: 304\AA\
({\em top left}), 171\AA\ ({\em top right}), 193\AA\ ({\em bottom left}),
211\AA\ ({\em bottom right}). The field-of-view is
180\arcsec$\times$132\arcsec}.
\label{fig:movie_aia}
\end{figure}

The movie of the evolution of the magnetic field is also provided using a larger
field-of-view to clearly see the supergranule
pattern. The movie is for the same period of eight hours and with a cadence of
45s.

\begin{figure}[!h]
\centering
\includegraphics[width=.8\linewidth]{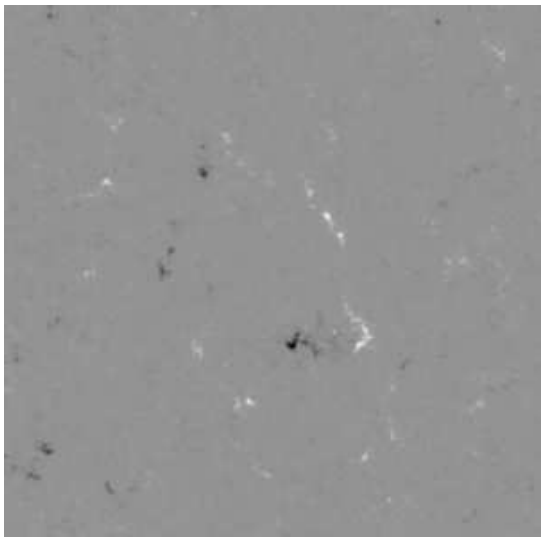}
\caption{Still image of the SDO/HMI line-of-sight magnetic field at 21:30 UT on
30 May 2010. The fiedl-of-view is 180\arcsec$\times$180\arcsec\ centered at
(-320\arcsec, -280\arcsec) at this particular time (reference frame for the
cross-correlation).}
\label{fig:movie_mag}
\end{figure}
 
\end{appendix}

\end{document}